\RequirePackage{fix-cm}
\documentclass[smallextended]{svjour3}  
\usepackage{algorithm}
\usepackage{algpseudocode}
\usepackage[utf8]{inputenc}
\usepackage{graphicx}
\usepackage{amsfonts}
\usepackage{booktabs}
\usepackage{tikz}
\usetikzlibrary{calc}
\usepackage{utopia}
\usetikzlibrary{arrows}
\usepackage{pgfplots}
\usepackage[utf8]{inputenc}
\usepackage{pgfgantt}
\usepackage{adjustbox}
 \usepackage{lscape}
 \usepackage{pgfgantt}
 \usepackage{arydshln}
\usepackage{graphicx}
\usepackage{xcolor}
\usepackage{adjustbox}
\smartqed  
\usepackage{graphicx}

\usepackage{multirow}

\usepackage{svg}
\usepackage{amsmath}
\usetikzlibrary{
    positioning,
    fit,
    calc,
    arrows.meta,
    decorations.markings,
    shapes.geometric
}
\definecolor{embeddings}{HTML}{daf0e2}
\definecolor{inv}{HTML}{b8e3c6}
\definecolor{attention}{HTML}{FFB347}
\definecolor{ffn}{HTML}{FF985A}
\definecolor{lang}{HTML}{84b5e0} 
\definecolor{norm}{HTML}{F2C894} 
\definecolor{task}{HTML}{b5d1eb} 

\definecolor{block}{HTML}{e3e1e1}
\definecolor{yes}{HTML}{d1e6c9}
\definecolor{yesborder}{HTML}{599b3e}
\definecolor{no}{HTML}{fadfc2}
\definecolor{noborder}{HTML}{e28e43}
\definecolor{clozeinput}{HTML}{ebf1f8}
\definecolor{clozeinputborder}{HTML}{2f61ce}
\definecolor{clozeoutput}{HTML}{e4c6d4}
\definecolor{clozeoutputborder}{HTML}{943766}

\makeatletter
\pgfdeclareshape{document}{
\inheritsavedanchors[from=rectangle] 
\inheritanchorborder[from=rectangle]
\inheritanchor[from=rectangle]{center}
\inheritanchor[from=rectangle]{north}
\inheritanchor[from=rectangle]{south}
\inheritanchor[from=rectangle]{west}
\inheritanchor[from=rectangle]{east}
\backgroundpath{
\southwest \pgf@xa=\pgf@x \pgf@ya=\pgf@y
\northeast \pgf@xb=\pgf@x \pgf@yb=\pgf@y
\pgf@xc=\pgf@xb \advance\pgf@xc by-10pt 
\pgf@yc=\pgf@yb \advance\pgf@yc by-10pt
\pgfpathmoveto{\pgfpoint{\pgf@xa}{\pgf@ya}}
\pgfpathlineto{\pgfpoint{\pgf@xa}{\pgf@yb}}
\pgfpathlineto{\pgfpoint{\pgf@xc}{\pgf@yb}}
\pgfpathlineto{\pgfpoint{\pgf@xb}{\pgf@yc}}
\pgfpathlineto{\pgfpoint{\pgf@xb}{\pgf@ya}}
\pgfpathclose
\pgfpathmoveto{\pgfpoint{\pgf@xc}{\pgf@yb}}
\pgfpathlineto{\pgfpoint{\pgf@xc}{\pgf@yc}}
\pgfpathlineto{\pgfpoint{\pgf@xb}{\pgf@yc}}
\pgfpathlineto{\pgfpoint{\pgf@xc}{\pgf@yc}}
}
}
\makeatother

\tikzstyle{doc}=[%
    draw,
    thick,
    align=center,
    color=black,
    shape=document,
    minimum width=20mm,
    minimum height=28.2mm,
    shape=document,
    inner sep=2ex,
]

\tikzset{
    embeddings/.style={
        black,
        draw=black,
        fill=embeddings,
        rounded corners=1mm, 
        minimum height=1cm
    }
}
\tikzset{
    inv/.style={
        black,
        draw=black,
        fill=inv,
        rounded corners=1mm, 
        minimum height=1cm
    }
}
\tikzset{
    attention/.style={
        black,
        draw=black,
        fill=attention,
        rounded corners=1mm, 
        minimum height=1cm
    }
}
\tikzset{
    ffn/.style={
        black,
        draw=black,
        fill=ffn,
        rounded corners=1mm, 
        minimum height=1cm, 
        
    }
}
\tikzset{
    norm/.style={
        black,
        draw=black,
        fill=norm,
        rounded corners=1mm, 
        minimum height=1cm, 
        
    }
}
\tikzset{
    lang/.style={
        black,
        draw=black,
        fill=lang,
        rounded corners=1mm, 
        minimum height=1cm, 
        
    }
}
\tikzset{
    task/.style={
        black,
        draw=black,
        fill=task,
        rounded corners=1mm, 
        minimum height=1cm, 
        
    }
}
\tikzset{
    fusion/.style={
        black,
        draw=black,
        fill=block,
        rounded corners=1mm, 
        minimum height=1cm, 
        
    }
}

\tikzset{
    yes/.style={
        black,
        draw=yesborder,
        fill=yes,
        rounded corners=1mm, 
        minimum height=2em, 
        
    }
}

\tikzset{
    no/.style={
        black,
        draw=noborder,
        fill=no,
        rounded corners=1mm, 
        minimum height=2em, 
        
    }
}

\tikzset{
    clozeinput/.style={
        black,
        draw=clozeinputborder,
        fill=clozeinput,
        rounded corners=1mm, 
        minimum height=2em, 
        
    }
}

\tikzset{
    clozeoutput/.style={
        black,
        draw=clozeoutputborder,
        fill=clozeoutput,
        rounded corners=1mm, 
        minimum height=2em, 
        
    }
}
\tikzset{
    ultra thin/.style= {line width=0.1pt},
    very thin/.style=  {line width=0.2pt},
    thin/.style=       {line width=0.4pt},
    semithick/.style=  {line width=0.6pt},
    thick/.style=      {line width=0.8pt},
    very thick/.style= {line width=1.2pt},
    ultra thick/.style={line width=1.6pt}
}

\tikzset{
    *|/.style={
        to path={
            (perpendicular cs: vertical line through={(\tikztostart)},
                                 horizontal line through={(\tikztotarget)})
            -- (\tikztotarget) \tikztonodes
        }
    }
}

\begin{document}

\title{Utilization of Pre-trained Language Models for Adapter-based Knowledge Transfer in Software Engineering\thanks{This research is supported by a grant from Natural Sciences and
Engineering Research Council of Canada RGPIN-2019-05175 and Mitacs Globalink award, 2021.}
}



\author{Iman Saberi       \and
        Fatemeh Fard          \and
        Fuxiang Chen
}


\date{Received: date / Accepted: date}

\institute{Iman Saberi \at
              University of British Columbia \\
              3333 University Way, Kelowna, BC V1V 1V7, Canada \\
              \email{iman.saberi@ubc.ca}           
           \and
           Fatemeh Fard \at
              University of British Columbia \\
              3333 University Way, Kelowna, BC V1V 1V7, Canada \\
              \email{fatemeh.fard@ubc.ca}           
           \and
           Fuxiang Chen \at
              University of Leicester \\
              University Rd, Leicester LE1 7RH, United Kingdom \\           
              \email{fuxiang.chen@leicester.ac.uk}           
}

\maketitle

\begin{abstract} 
  Software Engineering (SE) Pre-trained Language Models (PLMs), such as CodeBERT, are pre-trained on large code corpora, and their learned knowledge has shown success in transferring into downstream tasks (e.g., code clone detection) through the fine-tuning of PLMs. In Natural Language Processing (NLP), an alternative in transferring the knowledge of PLMs is explored through the use of \textit{adapter}, a compact and \textbf{parameter efficient} module that is inserted into a PLM. 
  Although the use of adapters has shown promising results in many NLP-based downstream tasks, their application and exploration in SE-based downstream tasks are limited. 
  
  Here, we study the knowledge transfer using adapters on multiple downstream tasks including cloze test, code clone detection, and code summarization.
  These adapters are trained on code corpora and are inserted into a PLM that is pre-trained on English corpora or code corpora. We called these PLMs as NL-PLM and C-PLM, respectively.
  We observed an improvement in results using NL-PLM over a PLM that does not have adapters, and this suggested that adapters can transfer and utilize useful knowledge from NL-PLM to SE tasks. The results are sometimes on par with or exceed the results of C-PLM; while being more efficient in terms of the number of parameters and training time. 
  Interestingly, adapters inserted into a C-PLM generally yield better results than a traditional fine-tuned C-PLM. Our results open new directions to build more compact models for SE tasks. 

\end{abstract}

\section{Introduction}
\label{sec:intro}

Pre-Trained Language Models (PLMs) such as BERT \cite{devlin2018bert} and RoBERTa \cite{liu2019roberta} are pre-trained on natural language text (i.e., sentences from Wikipedia). These PLMs provide rich linguistic representations and, when fine-tuned on multiple Natural Language Processing (NLP) downstream tasks such as text classification and language understanding \cite{liu2019roberta}, yield promising results.


In Software Engineering (SE), similar approaches were adopted -- pre-training a language model on code (\textit{instead of natural text}). In this work, we refer to the PLMs pre-trained on code and natural language text as \textbf{C-PLMs} and \textbf{NL-PLMs}, respectively. For example, CodeBERT \cite{feng2020codebert} and CuBERT \cite{kanade2020CuBERT} are two C-PLMs that were developed to obtain code representations. CodeBERT is a multilingual PLM pre-trained on code and the code comments, while CuBERT trains a BERT model using code. 
These C-PLMs are then fine-tuned on SE downstream tasks such as code clone detection and code search \cite{feng2020codebert,kanade2020CuBERT}.

Fine-tuning PLMs is the most common approach in transferring knowledge from existing models to downstream tasks. When a PLM is fine-tuned, \textit{all} of its learned weights are adjusted (i.e., \textit{relearned}) using the dataset of the downstream task. 
Although this approach achieves state-of-the-art performances on multiple NLP \cite{liu2019roberta,lan2019albert} and SE tasks \cite{feng2020codebert,kanade2020CuBERT,wang2021clsebert}, it is computationally expensive and space inefficient as all the PLM's parameters need to be relearned for every task, and each fine-tuned task need to be saved as another model. 
Therefore, researchers have been exploring and studying better ways to transfer knowledge from PLMs.

In NLP, \textit{adapter} has been introduced for the Transformer-based architecture. 
It is a parameter-efficient, compact, and more extensible approach to knowledge transfer \cite{houlsby2019parameterEfficient}. 
An adapter \textit{shares} the parameters of a PLM for \textit{all} of its tasks/languages while introducing a small set of task/language-specific parameters on the intermediate layers of a PLM. 
In this way, only a small set of weights are learned (\textit{as opposed to relearning all the weights of a PLM during the fine-tuning process}).
In recent years, a number of adapter-based architectures such as serial to parallel \cite{zhu2021serial}, language-specific transformations \cite{bapna-firat-2019-simple,artetxe-etal-2020-cross,philip2020language,zhu2021serial}, and task-specific transformations \cite{pfeiffer2020adapterfusion,pfeiffer2020madX} were proposed.
Even though adapters have shown promising results in the NLP domain, their capability has not been explored extensively in other domains, such as SE. Adapters have also not been studied to extend to other language modalities, such as programming languages. 
In addition, despite the similarities between programming languages and natural languages \cite{allamanis2018survey,hindle2016naturalness}, recent efforts on C-PLM are mostly on introducing new objectives, and there are limited studies to understand how knowledge is (\textit{can be}) transferred from natural languages to programming languages.

In this paper, we extended our previous work \cite{divyam} to explore the use of adapters for SE.
In our previous work, we studied the extent an adapter is used to transfer the representations of NL-PLMs. In this extended version, we further investigate adapters for knowledge transfer (\textit{using C-PLMs}) on SE-related tasks. 
This is first done through a Cross-Modal model, which we refer to as \textbf{MODE-X}. MODE-X trains adapters on code and inserts the trained adapters on the layers of NL-PLMs (our previous work). Then, we add adapters into C-PLMs to assess whether they can improve the results by transferring the learned knowledge of the same modality (i.e., code) in C-PLM using a small set of trainable parameters. 
The resulting models are a modified version of the original NL-PLMs or C-PLMs with adapters embedded in them. We evaluate the models on three common tasks: cloze test, code clone detection, and code summarization. 
These tasks evaluate neural models trained on code, and they are the evaluation benchmark tasks found in Cod-eXGLUE \cite{lu2021codexglue} \footnote{{https://github.com/microsoft/CodeXGLUE}}. 
We compare the results of our models (adapters on NL-PLMs and C-PLMs) with the results obtained by traditional fine-tuning of PLMs (i.e., NL-PLMs such as RoBERTa and C-PLMs such as CodeBERT and GraphCodeBERT). We also compare the number of learned parameters and the training time in the different models. 
Additionally, we apply attention and probing analysis to understand the learned representations using adapters.

Our results show several interesting phenomena:  1) When MODE-X trains adapters on RoBERTa (NL-PLM) for the code clone detection task, the resulting models yield better performance than just fine-tuning RoBERTa traditionally. When compared to the traditional fine-tuning of CodeBERT (C-PLM), the results are similar. 2) When MODE-X is used, or we train adapters on CodeBERT and GraphCodeBERT (C-PLMs) for the code summarization task, the majority of the results yield better performance than just fine-tuning the C/NL-PLMs traditionally. 3) Despite MODE-X having similar or better performance than fine-tuning C/NL-PLMs traditionally, the training time and the number of parameters are significantly lower.
We note here that the main objective of this work is not to present a new model architecture but to explore adapters in SE.

This is the first work that explores adapters comprehensively in SE. We empirically study adapters that are trained on (\textit{or where the knowledge is learned from}) NL-PLMs and C-PLMs for SE-related tasks. 
Compared to the previous version, we included the following items: i) updated the literature review section, ii) added a new task, code summarization, iii) evaluated the models with C-PLMs, and iv) analyzed the learned representations with probing tasks and attention analysis. In the new experiments, we are mainly interested in understanding the effect of adapters for other tasks and C-PLMs.



The rest of this paper is organized as follows. 
Section \ref{sec:LR} and \ref{sec:background} survey the existing literature and provide the background details on adapters, respectively.  The design of our study, including the description of our four research questions are described in Section \ref{sec:studyDesign}. For the research questions, their experimental setup and results (\textit{including discussions}) are shown in Section \ref{sec:rq1} (\textit{RQ1: Code Representation Using Adapters}), Section \ref{sec:rq2} (\textit{RQ2: Adapters' Ability for Cross-Modal Transfer to Code Clone Detection}), Section \ref{sec:code-summary} (\textit{RQ2 \& RQ3: Adapters' Ability for Code Summarization}) and Section \ref{sec:rq4} (\textit{RQ4: Computational Efficiency of Adapters}). Section \ref{sec:discussions} discusses our findings through probing and attention analysis of the adapters. We discuss the implications in Section \ref{sec:implications} and the threats to the validity of our work are discussed in Section \ref{sec:threats}. We then conclude the paper with future work in Section \ref{sec:conclusion-future}.

\section{Literature Review} \label{sec:LR}

Inspired by Transformers \cite{vaswani2017attention} and PLMs in NLP \cite{devlin2018bert,liu2019roberta,raffel2020t5,zhang2020pegasus}, several studies have emerged using Transformer-based PLMs for code representation \cite{kanade2020CuBERT,feng2020codebert,buratti2020cbert,tufano2020generating,roziere2021dobf,guo2020graphcodebert} in software engineering. CuBERT \cite{kanade2020CuBERT} and CodeBERT \cite{feng2020codebert} pioneered the pre-training of a BERT model \cite{devlin2018bert} for code. 
Consequently, C-BERT \cite{buratti2020cbert} and CodeTrans \cite{elnaggar2021codetrans}, based on the T5 architecture \cite{raffel2020t5}, were introduced.
Roziere et al. \cite{roziere2021dobf} present DOBF, an MLM-based pre-training objective that encourages code comprehension. 
The authors of CodeBERT \cite{feng2020codebert} were the first to incorporate bimodal pre-training for code; learning the NL-PL pairs from the CodeSearchNet corpora \cite{husain2020codesearchnet}. 
Concurrently, Tufano et al. \cite{tufano2020generating} showed that BART, a denoising autoencoder-based Transformer \cite{lewis2020bart}, initially pre-trained on a large English corpora and subsequently on a large corpus of code, can be fine-tuned for generating assert statements for unit tests. 
Drain et al. also used a pre-trained Transformer for generating bug fixes \cite{drain2021generating}.
CLSEBERT is developed and is used for four tasks, including code clone detection \cite{wang2021clsebert}. 
Although many different code-based PLMs were developed to represent code, 
they share a common property: they must be fine-tuned separately for each of the downstream tasks. This becomes an issue when scaling up the number of downstream tasks, as an entirely new model is required for every task.
Moreover, in multilingual PLMs like CodeBERT, the model learns features that can help its domain languages while discouraging representations that do not. Thus, it suffers from the ``curse of multilinguality'' when one begins to scale up the model to include new languages \cite{conneau2019curseofMultilinguality}.

NLP researchers have explored other avenues of efficient knowledge transfer to eliminate the shortcomings associated with the fine-tuning of large PLMs. 
The compact and extensible bottleneck layers, known as adapters, are one of the main techniques \cite{houlsby2019parameterEfficient} introduced. In terms of the learned model parameters, adapters use a small fraction of that of the original Transformer.

A number of adapter-based frameworks ranging from language-focused \cite{artetxe-etal-2020-cross,pfeiffer2020madX} to task-focused \cite{bapna-firat-2019-simple,pfeiffer2020adapterfusion} approaches were proposed. Bapna et al. \cite{bapna-firat-2019-simple} demonstrate the use of adapters in domain adaptation for Neural Machine Translation and employ them in a multilingual setting. 
Artetxe et al. \cite{artetxe-etal-2020-cross} transfer a monolingual PLM to an unseen natural language via adapters. Subsequent studies showed that it is promising to use multiple distinct task-specific adapters to disentangle different elements of knowledge relevant to the target domain of the downstream task \cite{pfeiffer2020adapterfusion} and that stacking task- and language-specific adapters is effective in adapting a multilingual model to an unseen natural language \cite{pfeiffer2020madX}.

Zeng et al. \cite{zeng2022extensive} conducted a comprehensive study on the existing NL-PLMs to enhance the understanding of their strengths and limitations. More specifically, they validated the performance of different NL-PLMs, compared such models with the previous domain-specific state-of-the-art models, and investigated the robustness of PLMs. They found that subtle performance fluctuations can refute the findings in the original papers, and none of the existing PLMs can dominate the other models. 
Shamil et al. \cite{ayupov2022parameter} evaluated the two approaches widely used for the parameter-efficient fine-tuning of transformers for code, namely adapters and LoRA. Other research through empirical studies showed that parameter efficient ﬁne-tuning approaches might outperform the traditional ﬁne-tuning method in NLP tasks with small data, and as the data size increases, the traditional ﬁne-tuning method could achieve better performance \cite{he2021towards,chen2022revisiting}.

\textbf{Differences of our work with the current literature:}
Although there are many studies on C-PLMs as well as in exploring adapters for NLP, there is no attempt to extend the adapters to other modalities, and few works have utilize adapters for programming languages and SE tasks, which we have explored in this paper. 
This paper extends our previous work, which is the first to assess adapters in SE \cite{divyam}. The closest study to ours is the work of Shamil et al. \cite{ayupov2022parameter} in which they perform two widely used parameter-efficient fine-tuning approaches for the C-PLMs. They have applied adapters for SE tasks and their main contribution is to evaluate the performance of adapters compared to the traditional fine-tuning approaches on C-PLMs (i.e., application of adapters as a quick fine-tuning approach). However, in this work, we mainly focus on highlighting the adapters' abilities to transfer knowledge from natural language to the programming language domain and then extend this ability to C-PLMs. 

\section{Background} \label{sec:background}


\subsection{Transformers and PLMs}
Transformers are the state-of-the-art neural network architecture that achieved promising results in multiple NL tasks \cite{vaswani2017attention}. 
Transformer consists of stacks of encoders and decoders layers. 
It uses an attention mechanism through the multi-head self-attention sub-layer, followed by a feed-forward sub-layer. 
The multi-head self-attention helps the model encode each word by attending to other words in the input sequence. 
Each of these sub-layers in the encoder has a residual connection, and a layer normalization is applied after each one (i.e., multi-head self-attention and feed-forward network). 
Bidirectional Encoder Representations From Transformers, BERT, is the base PLM for our study \cite{devlin2018bert}. 
BERT enables fine-tuning the model on downstream tasks with an additional output layer. 
After BERT, multiple Transformer-based PLMs were introduced. One popular PLM is RoBERTa \cite{liu2019roberta} and it is the main architecture for many C-PLMs.

\subsection{Adapters}

An adapter is introduced as a new layer (with an additional small number of parameters) that is inserted into a PLM to enable the PLM to adapt to a new language \cite{pfeiffer2020madX}. 
An adapter can be trained as either a language-specific adapter module (\textit{L-adapter}) or a task-specific adapter module (\textit{T-adapter}). The former is trained via the masked language modeling objective on an unlabelled target language dataset (this allows the PLM to adapt to unseen languages that are not covered in the PLM), and the latter is used to optimize a target task on a labeled dataset. 

\begin{figure}[h]
  \centering
  \includegraphics[width=\columnwidth]{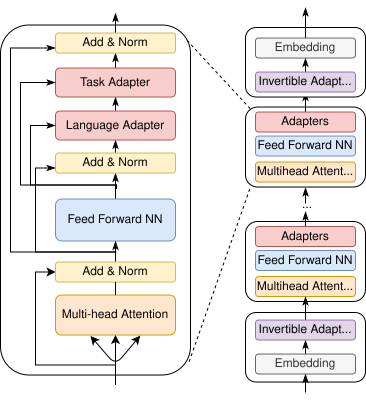}
  \caption{Language, task, and invertible adapters in the MAD-X framework, adapted from \cite{pfeiffer2020madX}.}
  \label{Adapters}
  
\end{figure}

The framework for adapters that we use in this study is based on the work 
\textit{Multiple Adapters for Cross-lingual transfer (MAD-X)} \cite{pfeiffer2020madX}, which uses architecture as its basis that allows the sharing of information between multiple tasks \cite{pfeiffer2020adapterfusion}. 
MAD-X enables the adaptation of unseen languages in a PLM ``without learning expensive language-specific token-level embeddings'' by freezing the initial learned parameters of the PLM and learning only a small number of parameters relative to that of the PLM.
The overall architecture of the adapters is shown in Figure \ref{Adapters}. 
The language and task adapter modules are inserted after the feed-forward network in \textit{each} layer of a Transformer-based PLM.
The T-adapters are stacked on top of the L-adapters when they are used for downstream tasks. 
The language adapter $LA_l$ at layer $l$ of the Transformer is defined as

\begin{equation}
    LA_l(h_l, r_l) = U_l(ReLU(D_l(h_l))) + r_l
\end{equation}

where 
$D \in {\mathbb{R}}^{h\times d}$, 
$h$ is the hidden size of the Transformer, $d$ is the dimension of the adapter, and $D$ is the down-projection. $ReLU$ is the activation function used and 
$U \in {\mathbb{R}}^{d\times h}$ 
is up-projection at every layer $l$. $h_l$ (output of the subsequent layer normalization) and $r_l$ (output of the feed-forward layer) are the hidden state and residual at layer $l$ of the Transformer, respectively.
During the training of T-adapters, the parameters of both the L-adapter and the Transformer are frozen. 
The task adapters, $TA_l$ at layer $l$ of the Transformer model is similar to $LA_l$ and is computed as below:

\begin{equation}
    TA_l(h_l, r_l) = U_l(ReLU(D_l(LA_l))) + r_l
\end{equation}

\textbf{Invertible Adapters}
\label{background:invertible_adapters}
The invertible adapters are proposed in \cite{pfeiffer2020madX} to deal with the mismatch between the vocabularies of a multilingual PLM and an unseen language. 
These are inserted on top of the input embedding layer, and their inverses are inserted before the output embedding layer, as shown in the left part of Figure \ref{Adapters}.
Each language of a multilingual PLM will have an invertible adapter. 
The function of the invertible adapters is similar to that of language adapters. It is used to capture the language-specific transformations at the token level. They are trained with language adapters using the MLM objective on a set of unlabeled dataset.
This inversion enables efficient utilization of the ``parameter budget''. This allows us to leverage the same set of parameters to adapt both the input and the output representations. 
For fine-tuning a model on a specific task, we remove the output embedding layer and its corresponding inversion layer, following which we freeze the parameters of the L-adapters and the PLM's parameters. 

\section{Study Design} \label{sec:studyDesign}

In this study, we seek the answers to the following research questions. 
 
\textbf{RQ1: How do adapters perform on code representation when adapting NL-PLM to a target programming language?}
    
    Each downstream task has its own programming language dataset. For example, for the code clone detection task, there are separate python-based and java-based datasets. We trained each \textit{language adapter} in the MODE-X architecture using the dataset of a programming language to learn the code representation. To study how well this code representation is being learned, we assess the language adapters using the cloze test task. \\

\textbf{RQ2: How effectively does MODE-X facilitate cross-modal (natural language to programming language) transfer on downstream tasks compared to the traditional approach of fine-tuning a PLM?}
    
    We study the impact that adapters of an NL-PLM e.g., ROBERTa, have on SE-related tasks (an adapter is trained on a single programming language/task). For example, we first train a task adapter on a SE-related task. This adapter is then inserted into RoBERTa and compared with other baseline models for the same SE-related task. 
    \\

\textbf{RQ3: Can adapters in C-PLM have better performance compared to the traditional approach of fine-tuning a C-PLM?}

    RQ3 is similar to RQ2, except that instead of training adapters on an NL-PLM e.g., ROBERTa, we use C-PLMs (\textit{e.g., CodeBERT and GraphCodeBERT}). Since the adapters are trained on SE-related tasks, we wanted to study if a higher performance can be achieved through training adapters on C-PLMs as compared to NL-PLMs.
    \\

\textbf{RQ4: How computationally efficient are adapters compared to the traditional approach of fine-tuning a PLM? }

    Fine-tuning a PLM on a downstream task generates another model that is similar to the PLM with respect to the number of parameters and the training time. 
    In this RQ, we are interested in studying the number of parameters needed, as well as the training time in each approach.


  

\subsection{Methodology Overview}

In this section, we first present an overview of our methodology and the details of each section are provided in Sections \ref{sec:rq1} to \ref{sec:code-summary}.
We first choose an NL-PLM as the base PLM. In our study, the chosen NL-PLM is RoBERTa, as it has a similar architecture to CodeBERT. 
We train three programming language (PL) specific adapters (i.e., L-adapters) on three sets of unlabeled dataset, one for C/C++, one for Java, and one for Python on the Masked Language Modeling objective. The L-adapters are inserted into each layer of RoBERTa.  
These three PLs are chosen based on the availability of the training and testing datasets on the downstream task.
We use the CodeSearchNet \cite{husain2020codesearchnet} dataset to train the L-adapters for Java and Python, and the CodeNET dataset \cite{puri2021projectCodeNet} to train the L-adapter for C/C++.

The tasks we chose for evaluation in our experiments are from the code-code category and code-natural language category in the CodeXGLUE benchmark \cite{lu2021codexglue} published by Microsoft. 
This benchmark is chosen as it is a popular collection of tasks, datasets, and platforms for evaluating and comparing machine learning models for code understanding and generation.
The two code-code tasks are Cloze Test (CT) and code clone detection. The third task (code-natural language) is code summarization.
These tasks are selected as they cover both PL-PL and PL-NL, thus allowing us to better understand the ability of adapters to transfer the learned knowledge of the PLMs to code-related tasks. 

Note that we work with both L-adapters (i.e., adapters trained in a programming language such as Java using MLM objective) and T-adapters (i.e., adapters trained on a specific task in one programming language such as code summarization in Java) in our work. For RQ1, only the L-adapters are used. For code clone detection, the L-adapters and T-adapters are used, and for code summarization, only the T-adapters are inserted within the layers of a PLM. The rationale for our choices and the reasons are illustrated in the subsequent paragraphs. 

Code clone detection is a task that mostly requires code only. So the code clone detection datasets do not have any code comments in natural language and consist of only code; therefore, it has only one modality.
Code summarization, on the other hand, consists of both code and natural language, therefore, having two modalities. 

We conduct experiments with slightly different approaches to evaluate the benefits of adapters in knowledge transfer from PLMs for each of these tasks. 
For code clone detection, we train both the language adapters and task adapters and insert them into RoBERTa. This helps the NL-PLM to learn both the PL and the code-specific task. 
For code summarization, we have opted to only train the task adapters in our approach. This decision was made given the similarities in architecture between the language and task adapters as well as the fact that code summarization involves both natural language and code.
These adapters are then inserted within the layers of PLMs to assess their performance. 

The rationale of choosing only task adapters for RQ3 is that the aim of a language adapter is to learn the language representation whereas the aim of a task adapter is to train an adapter for a specific target task such as code summarization. For adapting Roberta to a programming language target task, first, the model is required to learn the representations of the target language (e.g., python) by a language adapter, and then it needs to be trained on a target task using the task adapter. However, when we use C-PLMs, the model has already been trained on programming languages during the pre-training phase and so we do not need language adapters for C-PLMs. 
We specifically focus on the task of code summarization for our third research question (RQ3). This task is chosen due to its generative nature and the need to address both the NL and PL aspects of C-PLMs.

For RQ1, we evaluate the ability of the trained model to predict the missing tokens, i.e., Cloze Test (CT) task \cite{feng2020codebert}. This task shows how the contextual knowledge of the NL-PLM is transferred using adapters to a new modal, i.e., code \cite{lu2021codexglue}. 
Our model consisting of L-adapters inserted into the NL-PLM is tested for CT. 
The results of MODE-X are then compared to RoBERTa and CodeBERT.
The evaluation metric for cloze test is accuracy, as explained in Section \ref{sec:rq1}. Note that cloze test is only evaluated for the language adapters. 
 
For RQ2, code clone detection and code summarization are used. 
For code clone detection, we insert T-adapters on top of the L-adapters in all the layers within RoBERTa. 
For code summarization, only T-adapters are inserted within the layers of RoBERTa. 
Note that a task adapter is created for each task, i.e., code clone detection and code summarization. 
The parameters of the NL-PLM and L-adapters are frozen in the RQ2 experiments. It can be seen as another PLM with T-adapters injected into it to adapt the PLM's parameters away from its pre-trained MLM objective to a new objective on the downstream tasks.
We refer to the model containing adapters in the NL-PLM as \textbf{MODE-X}.
The results are evaluated against the fine-tuned NL-PLMs and C-PLMs on clone detection and code summarization. 
Three datasets are used for code clone detection in C/C++, Java, and Python with the evaluation metrics of F1 and MAP@R explained in Section \ref{sec:rq2}. 
Note that only the code clone detection task requires language adapters. The cloze tests in RQ1 are applied to these three languages.
For code summarization, smooth BLEU-4 is used as explained in Section \ref{sec:code-summary}, and we conduct the experiments on the six languages in the CodeSearchNet dataset, Ruby, JavaScript, Go, Python, Java, and PHP.

Based on the results we obtained for each of the tasks in RQ2, for RQ3, we consider the task of code summarization and add the related adapters to the C-PLMs, CodeBERT \cite{feng2020codebert} and GraphCodeBERT \cite{guo2020graphcodebert}. 
The main motivation for investigating code summarization is to study how to leverage the strengths of both NL and PL in the underlying network. We aim to evaluate the effectiveness of adapters in improving the results of C-PLMs in this context. Additionally, we aim to assess the performance of adapters on a generative task through this experiment.

Note that we choose CodeBERT and GraphCodeBERT to provide a fair comparison between RoBERTa (NL-PLM) and C-PLMs. These models are widely used in multiple software engineering studies. Additionally, the other C-PLMs are larger or have a different architecture, thus, threatening a fair comparison. So, we only use CodeBERT and GraphCodeBERT. In this manuscript, our aim is not to achieve state-of-the-arts performance in the selected tasks. Rather, we are studying the effect of adding adapters to the NL-PLM and C-PLMs. 

Finally, in RQ4, we compare the computational efficiency of PLMs (with adapters) with the traditional fine-tuned PLMs (without adapters).

\section{Code Representation Using Adapters}\label{sec:rq1}
To answer the first research question, we evaluate the ability of the language adapters to represent code using the cloze test task because cloze test evaluates the linguistic knowledge of the models.
In this section, we explain the experimental setup and then report the results of RQ1. 

\subsection{Experimental Setup}

\subsubsection{Dataset}
  
We used CodeSearchNet \cite{husain2020codesearchnet}, a joint dataset effort from GitHub and Microsoft Research that consists of code and code comment pairs in six programming languages to train the L-adapters. 

We selected three main benchmark datasets on code clone detection as one of the tasks to evaluate the performance of MODE-X, and one of them is in C/C++ language. However, CodeSearchNet does not include C/C++. Therefore, we chose the dataset from CodeNet to train the L-adapters for C/C++. The CodeNet dataset is a large-scale, high-quality dataset that is collected for studies on artificial intelligence for code, published by IBM \cite{puri2021projectCodeNet}.  
To train the L-adapters, we randomly split the dataset of each programming language into 90 (train): 10 (validation).
The L-adapters are trained on the training dataset and then evaluated on the validation dataset. 
The statistics of the datasets are shown in Table \ref{tab:datasetStats}.
Note that as only C/C++, Java, and Python language adapters are trained and used for code clone detection, the cloze test results are only depicted for these three programming languages. 

\begin{table}[]
    \caption{Statistics of datasets for training L-adapters. Each cell in table indicates the total number of samples at each category.}
    \label{tab:datasetStats}
    \centering
    \small
    \begin{tabular}{cccc}
         \textbf{Language}& \textbf{Train \#} & \textbf{Validation \#} & \textbf{Total \#}  \\
         \hline
         \multicolumn{4}{c}{\textbf{CodeNet (CN)}}\\
         \hline
         C/C++&  559,497 & 62,167 & 621,664\\
         \hline
         \multicolumn{4}{c}{\textbf{CodeSearchNet (CSN)}}\\
         \hline
         Java& 454,451 & 26,909  & 481,360 \\
         Python & 412,178 & 23,107  & 435,285 \\
         \hline
    \end{tabular}
\end{table}

\subsubsection{Task}

\textbf{Cloze Test (CT)} is a probing task that is designed by the authors of CodeBERT to evaluate their model's capability in learning the linguistic information of code without modifying the model's parameters \cite{feng2020codebert}. 

For the CT task, given a code snippet, the model will predict the masked token of interest. 
It has two setups: CT-all and CT-max/min. In the former, the model will predict tokens from the code, where the tokens are from the entire vocabulary, while in the latter, the tokens that are to be predicted by the model are from the \{max, min\} set. The CT-max/min setup evaluates the model's ability to understand code semantics \cite{lu2021codexglue}.
For testing a model on CT, no fine-tuning is required.
Both the CT-all and CT-max/min datasets are the combination of the validation and test sets of the CodeSearchNet dataset. 
{


CodeSearchNet does not include C/C++. We have tried to build such a dataset ourselves, but we were unable to find a dataset with similar vocabularies and thresholds to the CT task in CodeXGLUE.
Therefore, CT is applied only to the CodeSearchNet languages. 



\subsubsection{Training L-adapters}

We trained the L-adapters using the invertible configuration \cite{pfeiffer2020madX} as mentioned in Section \ref{background:invertible_adapters}. 
The L-adapters are trained on the code corpora of the dataset on each of the programming languages separately, leading to three L-adapters: 
C/C++-adapter, Java-adapter, and Python-adapter. 
The L-adapters are trained
using the Adam optimizer and a learning rate of 10E-4. 

Note that adapters do not handle out-of-vocabulary words, instead they adapt the model to have a proper distribution of words on a target domain.
The out-of-vocabulary words are handled by tokenizers. In the case of adapting a pre-trained natural language model to programming languages, we use language adapters to train the proper word distributions of a target language (e.g., Python) from a natural language word distribution (i.e., English). 

\subsubsection{Baselines} \label{subsec:baselines}
\textbf{RoBERTa} (Robustly optimized BERT approach) \cite{liu2019roberta} is based on BERT, and its difference from BERT is in its pre-training steps
-- it uses only the MLM objective and includes a longer input sequence. RoBERTa is used in previous SE studies \cite{zhang2020sentiment}, and it is the base model for many C-PLMs, including CodeBERT \cite{feng2020codebert}. 
RoBERTa is released in different model sizes, and we use the 12-layer architecture variant known as RoBERTa-base. 

\textbf{CodeBERT} is a BERT-based model trained on code \cite{feng2020codebert}. It is one of the models that is used as a baseline on the CodeXGLUE platform. 
CodeBERT uses the same architecture as RoBERTa-base and is trained on two objectives: MLM and Replaced Token Detection (RTD) \cite{Clark2020ELECTRA}. 
There are two publicly available versions of CodeBERT, one trained using the code corpus of CodeSearchNet dataset and utilized the MLM training objective, (CodeBE-RT\textsubscript{MLM}), while the other uses a combination of MLM and RTD objectives (CodeBERT) and it is trained on the bimodal dataset (i.e., code and documents) of CodeSearchNet. 
CodeBERT is trained on a combination of six PLs from CodeSearchNet. 
For the cloze test, we use CodeBERT\textsubscript{MLM}, as the cloze test task requires the model to predict the masked token. 
CodeBERT (i.e., the model trained on MLM and RTD) cannot perform cloze test as the final layers include a discriminator model. Thus, the CodeBERT authors only published the results of the MLM variant for the cloze test in their work \cite{feng2020codebert}.

These two models are chosen for comparison as they display the transferability of the adapters from NL to PL, where RoBERTa and CodeBERT are at two extremes. 
In addition, they both use a similar architecture, and this can provide a fair comparison between the models, especially for comparing their parameters and training time efficiency. 


\subsubsection{Evaluation Metric}

\textbf{Accuracy} is calculated as $\frac{TP+TN} {TP+TN+FP+FN}$. Here, TP refers to the number of records that are correctly identified as belonging to the positive class, whereas FP refers to the records that are incorrectly classified as belonging to the positive class.
TN are the records that are correctly predicted as negative examples, whereas FN are the records that are incorrectly predicted as belonging to the negative class.

\subsection{Results}


For this RQ, neither the L-adapters nor RoBERTa or CodeBERT are fine-tuned. We evaluate the trained models for the accuracy of the CT task, which is on evaluating the performance of L-adapters in capturing the representation of the PLs.
The results are presented in Table \ref{tab:clozeTest-Max/Min}. 
The RoBERTa\textsubscript{L-adapter} shows the model where the Python-adapter or the Java-adapter are trained and inserted into RoBERTa. 
The L-adapters are tested on CT on the PL that they were trained on. As mentioned earlier, in CT-All, the tokens to be predicted are from the entire vocabulary and in CT-max/min, they are from \{max, min\}. 


Note that RoBERTa is pre-trained on natural language and the L-adapters are used to adapt this NL-PLM to the PLs. 
We observe that with adapters the results fall between the results of RoBERTa and CodeBERT. For Java in CT-all, interestingly, Roberta\textsubscript{L-adapter} outperforms CodeBERT slightly.  
Note that our goal is not to improve the CodeBERT results. 
Instead, we evaluate whether by using adapters, we can transfer the knowledge from a NL-PLM. The results for CT show that the L-adapters can help improve the obtained results from RoBERTa, by infusing PL knowledge into it and leveraging the PLM's previously learned knowledge.

\begin{table}
  \caption{Accuracy scores of the models on Cloze Test (CT). Best scores are in bold and the second high scores are underlined. }
  \label{tab:clozeTest-Max/Min}
  \centering
  \small
  \begin{tabular}{ccc}
  \hline
    \textbf{Model}& Python & Java\\
    
    \hline
    \multicolumn{3}{c}{\textbf{CT-max/min}}\\
    \hline
    \textbf{RoBERTa} & 59.18	& 59.75	\\
    \textbf{RoBERTa\textsubscript{L-adapter}} & \underline{66.30}&	\underline{66.81}\\
    
    \textbf{CodeBERT\textsubscript{MLM}} & \textbf{79.27}& \textbf{91.08}\\
    \hline
    \multicolumn{3}{c}{\textbf{CT-all}}\\
    \hline
    \textbf{RoBERTa} & 54.49 & 50.75\\
    \textbf{RoBERTa\textsubscript{L-adapter}} & \underline{74.35}& \textbf{75.63}\\
    
    \textbf{CodeBERT\textsubscript{MLM}} & \textbf{83.33}&	\underline{75.53}\\
    \bottomrule
  \end{tabular}
\end{table}

\section{Code Clone Detection Adapters}\label{sec:rq2}

In RQ2, we are interested in evaluating whether MODE-X can facilitate the cross-modal transfer from natural language to programming language. For this purpose, two tasks are chosen. The first is the code clone detection task that will be discussed in this section, and the second is the code summarization task in the next section (Section \ref{sec:code-summary}). 
In this section, we first explain the experimental setup before detailing the results. 

\subsection{Experimental Setup}

\subsubsection{Dataset}

We conducted experiments on code clone detection in three programming languages: C/C++, Java, and Python. For C/C++, the L- adapters are trained using the CodeNet dataset \cite{puri2021projectCodeNet}.
For Java and Python, the L-adapters are trained using the CodeSearchNet \cite{husain2019codesearchnet} dataset. 
We utilize the POJ-104 and the Big Clone Bench (BCB) datasets which are a part of the code-code pipeline dataset from CodeXGLUE \cite{lu2021codexglue}. 
POJ-104 contains C/C++ programs, and it is used for retrieving the top-k semantically similar code fragments. The retrieval result is then evaluated using the MAP@R score (see below for details). BCB is used to discern whether a pair of code fragments are semantically equivalent, and it is evaluated using the F1 score (see below for details). 
As there is no Python dataset in CodeXGLUE to be used for the code clone detection task, we consider the python-specific subset from the cross-language clone detection (XCD) dataset  \cite{perez2019cross}. We refer to this as the SCD-88 dataset, where 88 refers to the number of problems with multiple submitted solutions on Python.
We then reformulate it as a retrieval task and evaluate using the MAP@R score, similar to POJ-104. 
Table \ref{tab:ccdsplits} shows the respective splits for POJ-104, BCB, and SCD-88.

\begin{table}[]
    \caption{Statistics of code clone detection datasets}
    \label{tab:ccdsplits}
    \centering
    \small
    \begin{tabular}{cccc}
         \textbf{Dataset}& \textbf{Train \#} & \textbf{Validation \#} & \textbf{Test \#}  \\
         \hline
          \textbf{POJ-104} &  32,000 & 8,000 & 12,000\\
         \hline
         \textbf{BCB} &  901,028 & 415,416 & 415,416\\
         \hline
         \textbf{SCD-88} &  7,800 & 1,040 & 2,600\\
         \hline
    \end{tabular}
    \vspace{-2.5mm}
\end{table}

\subsubsection{Task}

\textbf{Code Clone Detection (CCD)} 
involves the identification of code fragments that share similarities within a given codebase, enabling developers to manage and maintain the code more efficiently. The primary objective of CCD is to accurately locate these similar code fragments and group them together to avoid redundancy, inconsistencies, and other potential issues that can arise from duplicated code \cite{wang2021clsebert}. 

Overall, the ability to effectively detect and manage code clones is essential in maintaining software quality, reducing development time, and improving productivity. With the help of CCD and C-PLMs, developers can optimize their codebase, improve their workflows, and deliver high-quality software products.

\subsubsection{Training Models}

\textbf{Training the T-Adapters:} We follow the same configuration as Jonas et al. to train the T-adapters \cite{pfeiffer2020adapterhub}. 
We use in-batch negative sampling to train these adapters while keeping in line with the experimental setup described by the authors of CodeBERT \cite{feng2020codebert}. To prevent the adapters from overfitting, dropout and early stopping are used. 



\textbf{Training the Baselines:} To maintain consistency across our evaluations, we re-trained and re-evaluated the existing downstream task benchmark performances of RoBERTa and CodeBERT in our study. 
The authors of CodeBERT have confirmed that our obtained results are accurate, although our results fall within 2\% error rate of what was reported in CodeBERT.
Keeping in line with the benchmark experiments of CodeXGLUE, we also utilize in-batch negative sampling. The choice of hyperparameters, learning rate schedules, and optimizers was similar to CodeXGLUE's benchmarking experiments.

All the experiments were conducted on an Nvidia Tesla V100 32GB GPU.

\subsubsection{Baselines}


Similar to cloze test, we compare the results with RoBERTa and CodeBERT. For code clone detection, we use the full CodeBERT, i.e., the model that is trained on MLM+RTD objectives. 

\subsubsection{Evaluation Metric} 

\textbf{\textit{F1-Score (F1):}} F1 Score is the weighted average of Precision and Recall: $F1 = \frac{2 \cdot (P \cdot R)}{P + R}$. 
Here, P stands for Precision, and it is computed as $P = \frac{TP}{TP+FP}$, whereas R is the Recall and it is calculated as $R = \frac{TP}{TP+FN}$. \\

\textbf{Mean Average Precision at R (MAP@R)} \cite{musgrave2020metric} is a metric used for measuring the informative accuracy, which mitigates the weakness of the R-Precision metric and only accounts for the ranking of the correct retrievals. In R-Precision, a score of $r/R$ is assigned to each query, wherein each query (e.g., a code that we want to find similar code samples from), we find the r nearest samples that are in the same class as the query from a total number of references, R. Here, R denotes the total number of references in the searchable dataset. 
Therefore, MAP@R calculates the Mean Average Precision based on the number of nearest neighbors for each sample, relative to R. For a single query, it is defined as follows where $P(i)$ is the Precision at $i$ if the $i$th retrieval is correct and $0$ otherwise: 
\[MAP@R = \frac{1}{R} \sum_{i=1}^{R} {P(i)}\]

\subsection{Results}

The results of MODE-X for code clone detection are shown in Table \ref{tab:CCD}. 
The programming language of the adapters in MODE-X is shown as subscript. 
For the T-adapters in natural language, it is reported that the last layer of the model learns the MLM objective better \cite{pfeiffer2020unks} -- better results are obtained when the L-adapters are dropped from the last layer, leaving only the T-adapters in this layer. 
In this RQ, we ran the following ablation experiments i) when we did not drop the L-adapters, ii) when dropping the L-adapters from the last layer (layer 12), and iii) when dropping the L-adapters from the last two layers (layers 11 and 12). We reported the best scores obtained, though the difference among the scores in each experiment was very low. 
Similar to NLP T-adapters, the best results were obtained when the L-adapters are dropped from the last or the last two layers.
For BCB and SCD-88, the best scores are from the model with the dropped L-adapter from its final layer.
The best results achieved for POJ-104 is by dropping the L-adapters from the last two layers.

For all three datasets, the results of MODE-X are between the results of the fine-tuned RoBERTa and CodeBERT on the code clone detection task. 
For the C/C++ and Python datasets, the adapters' results are 3-4 MAP@R points below CodeBERT. 
An interesting observation is that CodeBERT is not pre-trained on C/C++, but on other programming languages. It is only fine-tuned on C/C++ for the code clone detection task. The higher score of CodeBERT, in this case, is related to its learned knowledge from the other programming languages. In comparison, RoBERTa has not seen any programming language during pre-training. However, adding the C/C++-adapters to its layers helps to improve the model's results for code clone detection, which is similar to CodeBERT's results. 
For Java language, MODE-X improves the results of RoBERTa and has similar scores with CodeBERT. 
Note that Java is among the programming languages that CodeBERT is pre-trained and fine-tuned on. 

\begin{table}
  \centering
  \caption{Scores of the code clone detection for RoBERTa, CodeBERT, and MODE-X. The best scores are bold. }
  \label{tab:CCD}
  \small
  \begin{tabular}{ccc}
    \toprule
    Model & Dataset & Score\\
    
    \hline
    \textbf{RoBERTa} & POJ-104 & 81.52 (MAP@R) \\
    \textbf{MODE-X\textsubscript{C/C++}} & POJ-104 & {82.40} (MAP@R)\\
    \textbf{CodeBERT} & POJ-104 & \textbf{86.48} (MAP@R) \\
    
    \hline
    \textbf{RoBERTa} & BCB & 95.96 (F1) \\
    \textbf{MODE-X\textsubscript{Java}} & BCB & {96.61} (F1)\\
    \textbf{CodeBERT} & BCB & \textbf{96.65} (F1) \\
    
    \hline
    \textbf{RoBERTa} & SCD-88 & 73.90 (MAP@R) \\
    \textbf{MODE-X\textsubscript{Python}} & SCD-88 & {75.65} (MAP@R)\\
    \textbf{CodeBERT} & SCD-88 & \textbf{78.95} (MAP@R) \\
    \hline
  \end{tabular}
\end{table}


In this study, we focus on evaluating the effectiveness of the cross-modal transfer abilities of the adapters. Therefore, we train the adapters solely on the MLM objective, whereas CodeBERT is trained on two objectives of MLM and RTD. The RTD explicitly injects the code information into the CodeBERT's representation space. Although the impact of this dual objective may be unclear for code clone detection, CodeBERT has been reported to have better results over CodeBERT\textsubscript{MLM} for multiple tasks \cite{feng2020codebert}. 
The MODE-X results are close to CodeBERT while being more parameter efficient. 

We note here that for training the T-adapters, we used the recommended hyperparameters on AdapterHub \cite{pfeiffer2020adapterhub}. 
We also ran additional experiments with different learning rates for code clone detection on the SCD-88 dataset. When a different learning rate is used ($5E-4$), the results of MODE-X are improved to 79 MAP@R, which is equal to the CodeBERT's scores.  
Finding the best hyperparameters for adapters is not the scope of this work, but it is worth noting that adapter results may be improved further when the best hyperparameters are chosen.

\section{Code Summarization Adapters} \label{sec:code-summary}

In the code clone detection experiments, we observed that adapters aid in transferring the learned knowledge from an NL-PLM. 
In the second task, we apply adapters to code summarization to evaluate the knowledge transfer using adapters. For code summarization, we also take one step further to see whether adapters can boost the performance of C-PLMs. These are the questions asked in RQ2 and RQ3 that we seek the answers to in this section. 
Note that the results of Roberta are close to CodeBERT for code clone detection (see Table~\ref{tab:CCD}). Thus, we only apply code summarization in the rest of the experiments. Code summarization is a more challenging task as it is a generative task, and it includes the use of both code and its comments. 

\subsection{Experimental Setup}

\subsubsection{Dataset}
We used the CodeSearchNet \cite{husain2019codesearchnet} dataset, which comprises bimodal data, the code, and their corresponding comments, for six programming languages: Ruby, Javascript, Go, Python, Java, and PHP. 
For code summarization, we train the T-adapters separately for all these six programming languages. 
The statistics of the dataset are shown in Table \ref{table:csn-stat}. Note that we only use the bimodal part of the data as code summarization involves both the code and its comments.

\begin{table}[h!]
\centering
\caption{CodeSearchNet dataset for code summarization. Each cell in table indicates the number of samples for each programming language. }

\begin{tabular}{|c | c | c|} 
 \hline
 Language & bimodal Data  \\ [0.5ex] 
 \hline
 Ruby  & 52,905  \\
 \hdashline
 JavaScript  & 143,252  \\
 Go  & 317,832   \\ 
 \hdashline
 Python  & 458,219  \\ 
 Java  & 500,754  \\ 
 PHP  & 662,907 \\ 
 \hline
\end{tabular}

\label{table:csn-stat}
\end{table}

\subsubsection{Task}

\textbf{Code summarization} is a common SE task that generates a description of a given code \cite{chen2022transferability}. 
This is considered a PL-NL downstream task, as the input to the model is the code, and the output is a generated text in natural language.

\subsubsection{Training Models}
As mentioned in Section \ref{sec:studyDesign}, for code summarization, we only use T-adapters. 
For each baseline model, we inserted the T-adapters into the PLM, fixed the PLM weights, and then trained them for each programming language on the code summarization task.
T-adapters are trained on a 4x NVIDIA V100 GPU configuration, with a total of 30,000 training steps. A batch size of $32$ is utilized, along with the AdamW optimizer and a learning rate of $10E-5$.

\subsubsection{Baselines}

We consider \textbf{RoBERTa} \cite{liu2019roberta} as our baseline model, which is pre-trained on a large corpus of English data with a mask language modeling objective function. 
Similar to the code clone detection task, we also compare the results with that of CodeBERT.
Additionally, to answer RQ3, we use CodeBERT and GraphCodeBERT as the C-PLMs where we explore the performance of adding adapters into a C-PLM.
We compare the results between the fine-tuned C-PLMs and C-PLMs with adapters.
Details of RoBERTa and CodeBERT are found in Section \ref{subsec:baselines}. Here, we only provide the description for GraphCodeBERT. 

\textbf{GraphCodeBERT} \cite{guo2020graphcodebert} uses BERT as its backbone and includes the data flow of code to pre-train on three objective functions, Mask Language Modelling, Edge Prediction (in which some edges on the data flow graph are masked, and the goal is to predict the correct value for them), and Node Alignment (in which the model is required to predict the correct alignment between code and data flow).

\subsubsection{Evaluation Metric}
We used the smoothed BLEU-4 \cite{papineni2002bleu} score to evaluate the performance of the model on generating document summaries.
BLEU, Bilingual Evaluation Understudy, is used to count the proportion of n-grams in candidate text that appear in the ground truth code description, where $n$ can be $1,2,3$ or $4$. The higher the BLEU value, the higher the quality of code summarization is \cite{zhang2022survey}.
In this study, we utilize the smoothed BLEU-4 variant \cite{lin2004orange} by adding a small constant to the numerator and denominator of the precision calculation, allowing for a more lenient evaluation of the generated summaries. To calculate the BLEU score, we first measure the brevity penalty factor (BP) as follows:

\begin{equation}
  BP =
    \begin{cases}
      1 & \text{if $c > r$}\\
      e^{1-r/c} & \text{if $c \leq r$ }\\
    \end{cases}       
\end{equation}

where $c$ is the length of the candidate translation and $r$ is the effective reference corpus length. Then, we take the geometric mean of the test corpus’ modified precision scores, and multiply the result by the brevity penalty factor ($BP$) as follows:

\begin{equation}
  BLEU = BP.exp \big( \sum_{n=1}^N w_n \text{log} P_n \big)   
\end{equation}
Here, $P_{n}$ is the geometric average of the n-gram precision, using n-grams up to length $N$ and $w_{n}$ is the positive weights summing to one.

\subsection{Results}

\textit{Transferability of MODE-X from natural language to code (RQ2):} 
To evaluate the extent to which we could adapt an NL-PLM on a PL-NL target task, we plug the T-adapter into RoBERTa and train them on the target task for each programming language separately.  
As shown in Table \ref{table:different_methods}, MODE-X outperforms the fine-tuned CodeBERT on Ruby, Go, and Java and has similar results with the fine-tuned CodeBERT for the other three languages. 
In both the code clone detection and code summarization tasks, MODE-X improved on the results of RoBERTa, though we got better results for code summarization, and MODE-X \textit{outperforms} the fine-tuned CodeBERT's results for three languages. 
Note that MODE-X is RoBERTa plus task adapters, meaning that the underlying PLM is only trained on natural language, and the only introduction of code into the PLM is through the task adapters. 
Moreover, in MODE-X, the number of trainable parameters is much less than in CodeBERT. In CodeBERT, we fine-tune the entire model, which is pre-trained on programming languages. However, we can still outperform it by transferring from a natural language model.

\textit{Ability of adapters when a C-PLM is used (RQ3): }
Similar to the code clone detection task in Section \ref{sec:rq2}, we obtained better results on MODE-X for the code summarization task. 
In RQ3, we investigate to what extent we can improve the code summarization results of C-PLMs such as CodeBERT by inserting adapters.
To answer this question, we chose two widely used C-PLMs, CodeBERT and GraphCodeBERT. We insert adapters in their layers and then train the models as we did for RoBERTa. In Table \ref{table:different_methods} the results of these models are shown as CodeBERT\textsubscript{T-adapter} and GraphCodeBERT\textsubscript{T-adapter}. 
These results are compared with the scores of CodeBERT and GraphCodeBERT, which are fully fine-tuned on the code summarization datasets. 
As shown in Table \ref{table:different_methods}, when adapters are inserted in these models, the majority of the results are improved (except for Java in CodeBERT). 
For all languages except Python, the results of CodeBERT\textsubscript{T-adapter} are higher than the fine-tuned CodeBERT. Similarly, the results of the GraphCodeBERT\textsubscript{T-adapter} are higher than the fine-tuned GraphCodeBERT for Ruby, JavaScript, Go, and Python, and we observed similar results for Java and PHP. 

We relate this improvement in results to the fact that adapter-based fine-tuning can enhance the performance and stability of models during the fine-tuning phase \cite{he2021effectiveness}.
He et al. \cite{he2021effectiveness} conducted a comprehensive study on the effectiveness of adapters, focusing on the accuracy of the models rather than the efficiency of parameters. They argue that adapter-based fine-tuning can produce superior results compared to fully fine-tuning in low-resource languages, i.e., the languages for which small training data is available. Furthermore, Lee et al. \cite{lee2019mixout} proposed a regularization method called mix out to encourage the PLM's weights to remain closer to their initial pre-trained values during fine-tuning. As adapters kept the PLM's weights fixed during fine-tuning, adapter-based fine-tuning can enhance performance.

We summarize the three major findings as seen in Table~\ref{table:different_methods}: i) MODE-X has improved or on-par results than/with C-PLMs, ii) C-PLMs with adapters achieve improved results compared to when they are fully fine-tuned for most languages (compare each model with and without adapters, e.g., Code-BERT\textsubscript{T-Adapter} with CodeBERT), and iii) when we add adapters to the C-PLMs, higher scores are obtained for Ruby, JavaScript, and Go. These three languages have fewer records in the dataset, and it shows the effect of using adapters when our target language has fewer records.
Studying the advantages of using adapters for these languages is out of scope in our study.  
These findings show the importance of using adapters compared to fully fine-tuning the C-PLMs for code summarization.


\begin{table*}[t!]
\centering
\caption{Smooth BLEU-4 scores on code summarization. CodeBERT\textsubscript{T-Adapter} is fine-tuned on the monolingual dataset for each programming language (same as CodeBERT). The best scores are bold, and the second-best scores for each language are underlined. Note that MODE-X results are better than or on par with CodeBERT results. Moreover, when we add adapters to CodeBERT or GraphCodeBERT, the results are better than the C-PLMs.
This demonstrates that if we encourage the model weights to be closer to the PLM, we could improve the fine-tuning results without using additional data.}
\begin{tabular}{l | c | c| c| c| c| c} 
 \hline
 \textbf{Models} & \textbf{Ruby} & \textbf{JavaScript} & \textbf{Go} & \textbf{Python} & \textbf{Java} & \textbf{PHP} \\ [0.5ex] 
 \hline

 GraphCodeBERT\textsubscript{T-Adapter} & \textbf{14.53} & \textbf{16.54} & \textbf{23.74} & \underline{18.73} & \underline{19.08 } & 25.05 \\ 
 CodeBERT\textsubscript{T-Adapter} & \underline{14.12} & \underline{15.67} & \underline{23.21} & 18.47 & 18.99 & \textbf{25.55} \\ 
 MODE-X & 12.79 & 14.20 & 23.05 & 17.72 & 18.43 & 24.27 \\
\hline
 
 GraphCodeBERT   & 12.62 & 14.79 &18.40 & 18.02 & \textbf{19.22} & \underline{25.45}\\ 
 CodeBERT  & 12.16 & 14.90 &18.07 & \textbf{19.06} & 17.65 & 25.16 \\ 
 RoBERTa \cite{liu2019roberta}  & 11.17 & 11.90 &17.72 & 18.14 & 16.47 & 24.02 \\ 
 \hline
\end{tabular}

\label{table:different_methods}
\end{table*}



\section{Computational Efficiency of Adapters} \label{sec:rq4}

  

We evaluate the efficiency of the adapters based on their parameter budget and training time, 
with that of a traditional fine-tuned PLM. 
\emph{The parameter budget is the number of learnable parameters in the model.} For adapters, as we do not re-train RoBERTa, the parameter budget is the number of parameters required for training the adapters only.
We report 
the parameter budgets of the adapters for the entire 12 layers of the model.




CodeBERT and GraphCodeBERT have $\sim110$ million of parameters, and RoBERTa has $123$ million parameters. The parameter budget for task adapters has $\sim0.9$ million parameters. For code summarization, a decoder stack for the code summary generation is trained from scratch during the adapter fine-tuning process, which allocates $\sim47.8$ million parameters. In total, the number of trainable parameters for code summarization is around $\sim48.7$ million parameters during the adapter-based fine-tuning phase.
The L-adapters have a parameter budget of $7.39$. 
For code clone detection on POJ-104 and SCD-88, the number of parameters (in millions) required for MODE-X is $8.29$ ($7.39$ for L-adapters + $0.9$ for T-adapters). 
For code clone detection on the BCB dataset, MODE-X requires more parameters, a total of $9.46$ million parameters. The difference is that the code clone detection problem is a retrieval-based one for the other two and a classification problem for BCB. 


For task-specific fine-tuning, we only consider the parameters that are required for fine-tuning CodeBERT and the parameters for training T-adapt-ers in MODE-X (i.e., excluding the pre-training parameters of CodeBERT and the parameters of L-adapters).
For task-specific fine-tuning, adapters are $53 = 110/(9.46-7.39)$ times and $122.2 = 110/0.9$ times more parameter efficient than CodeBERT on BCB, and POJ-104/SCD-88, respectively.
For code summarization, we also obtain the same results of $122.2$ times more efficiently.
When considering the overall budget, i.e., the number of parameters required for training and fine-tuning CodeBERT and the number of parameters used for training L-adapters and T-adapters are $11.63 = 110/9.46$ to $13.27 = 110/8.29$ times more efficient than CodeBERT for code clone detection, and $2.25 = 110/48.7$ times more efficient for code summarization. Note that this efficiency is lower for code summarization as we need to add a decoder for this task in addition to the adapters.

It is worth mentioning that pre-training the CodeBERT needs $384$ hours of training  on 16 interconnected V100s for a batch size of $64$ \cite{feng2020codebert}. In contrast, L-adapters need $35$ hours of training on a single V100 GPU for the same batch size.  
Moreover, fine-tuning CodeBERT required between $2.5$ hours and an hour for T-adapters for code clone detection.
As adapters are significantly more parameter efficient, they have shorter training and inference time and, hence are more suitable in practice. 
The task of training adapters for code summarization requires approximately $10$ hours per programming language, utilizing $30,000$ training steps and consuming a similar parameter budget to T-adapters used for code clone detection. Note that the training time depends on the number of training steps -- since we train all the programming languages with the same number of steps, they have the same training time.


\section{Discussions} \label{sec:discussions}

In this section, we provide additional insights into our findings through two experiments, analysis of the probes, and attention of the models. The probing analysis unveils the characteristics of the learned code embeddings in the models. It reveals their ability for code surface, structure, and semantics information, through the three probing tasks we study here: Code Length Prediction, Cyclomatic Complexity, and Abstract Syntax Tree (AST) Node Tagging. The second experiment on attention analysis shows qualitatively how the attention in different layers is changed using adapters. This could furnish our understanding of the results in previous sections. 

\subsection{Probing Analysis} \label{sec:probing-analysis}
One way to investigate if programming language models encode the characteristics of code well enough to be applicable to a broad spectrum of downstream tasks is with the diagnostic task known as probing \cite{karmakar2021pre}.
Probing is used to determine if specific properties of the input can be predicted based on the fixed, pre-trained vector embeddings of a model. The effectiveness of probing in predicting these properties suggests whether or not the desired information is present in the model. Probing has been extensively studied in natural language models \cite{adi2016fine,alain2016understanding,peters2018dissecting,belinkov2017neural}.
It consists of two main components: a probing task and a probing classifier. The probing task is a diagnostic task designed to determine whether a specific property is encoded in the PLM's weights. 
The probing classifier is trained on the probing task using input vectors extracted from the frozen hidden layers of the PLM. It is usually a linear classifier with no hidden layers of its own. If the probing classifier can accurately predict a particular attribute from the pre-trained embeddings, we could infer that the original model has encoded this information in its hidden layers. The raw accuracy of probing is not the primary focus of the analysis, but rather probing is used to compare the encoding of the characteristic between different models or layers within a model \cite{karmakar2021pre}.

We evaluate the syntactic, structural, and semantic characteristics of {C-PLMs} and NL-PLM with adapters to assess how they could be adapted for programming languages.
We construct three initial probing tasks on RoBERTa \cite{liu2019roberta}, CodeBERT \cite{feng2020codebert} and GraphCodeBERT \cite{guo2020graphcodebert} with and without L-adapters: Code Length Prediction, Cyclomatic Complexity Prediction, and Abstract Syntax Tree Node Tagging.

\textbf{Code Length Prediction (LEN)}
The Code Length Prediction (LEN) task is a method for evaluating the ability of language models to encode surface-level information within code snippets \cite{karmakar2021pre}. Our hypothesis is that the length of a code snippet, when presented as a sequence of code tokens, should be relatively straightforward to predict for transformer models. To evaluate that, we use the Java dataset proposed by Karmakar and Robbes \cite{karmakar2021pre}. Each sample in the dataset consists of a code snippet and a label indicating its length. The probing task is to predict the label of each sample correctly.

\textbf{The Cyclomatic Complexity (CPX) }
The Cyclomatic Complexity (CPX) task is an evaluation method designed to investigate the ability of the code models to encode structural information within code. Cyclomatic complexity is an intrinsic characteristic of any code, and thus, it should be predictable for models without the need for explicit fine-tuning \cite{karmakar2021pre}. Predicting the complexity of code based solely on the sequence of tokens rather than understanding the underlying control flow is a challenging task. This is because it requires a deep understanding of the code and the ability to infer the relationships between different parts of the code. This is particularly difficult for programming languages that have complex control flow structures, such as loops and branches.

\textbf{Abstract Syntax Tree (AST) Node Tagging }
Abstract Syntax Trees (AST) consist of rich syntactical information and they are the bases for many structural code representation approaches \cite{karmakar2021pre,allamanis2017learning,brockschmidt2018generative}. One way to evaluate whether a PLM is good at code tasks such as code clone detection is that it should be able to learn and interpret the syntactic information of a sequence of code tokens as well as to predict them  correctly. The authors from \cite{karmakar2021pre} provided a Java dataset consisting of the Java code tokens and their corresponding node tags extracted from the AST, which we use in our analysis.

These three tasks are meant to assess whether the models are able to capture syntactic, structural, and semantic features of code, respectively. The tasks were chosen to cover the most commonly identiﬁable abstractions of code.
Note that the objective of the probing task here is to assess the ability of adapters to adapt NL-PLMs to code-related target tasks, as compared to C-PLMs. To this end, we will compare the results of the probing task conducted on RoBERTa with L-adapters, CodeBERT, and GraphCodeBERT. 
Additionally, we are interested in analyzing the effect of L-adapters on C-PLMs. Therefore, for each probing task (shown in Figures \ref{fig:len-prob}--\ref{fig:type-probe} and explained in detail in the next sub-sections), we have six plots, indicating the results for each of the three models with and without L-adapters. In each plot, the performance of the models is assessed layer-wise, where the performance is demonstrated on the y-axis, and the x-axis represents the layer number. 


\begin{figure}

\includegraphics[width=\columnwidth]{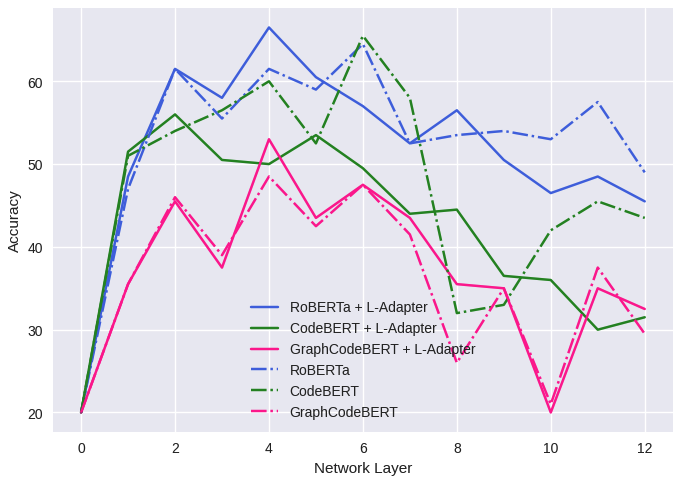}
\centering
\caption{The accuracy of predicting code length in the probing task is evaluated on each layer of the models. The x-axis represents the classification results, with the first layer being the input embeddings, serving as the naive baseline accuracy. All models have 12 Transformer layers, and the y-axis displays the accuracy at each layer for each model.}
\label{fig:len-prob}
\end{figure}

\subsubsection{Probing Analysis Results}

The analysis in this subsection is conducted to determine the behavior of NL-PLMs and C-PLMs with/without language adapters. 
In Figures \ref{fig:len-prob}, \ref{fig:cpx-prob}, and \ref{fig:type-probe}, the solid lines show the results of the models with L-adapters, and the dashed lines are the original models (i.e., without L-adapters). For easier comprehension, the colors of the plots are associated with the models.

{\textbf{Code Length Prediction Results.}
Figure \ref{fig:len-prob} demonstrates the classification results of the code length prediction task. 
The performance of RoBERTa with adapters is found to be comparable to that of the two C-PLMs without adapters. 
As depicted in the figure, the prediction of code length, is primarily determined by the lower layers of the models (e.g., the fourth layer), and the accuracy is decreased in the higher layers. In layer four, RoBERTa+L-adapter achieves the highest accuracy among all models.

Two interesting observations in Figure \ref{fig:len-prob} are that L-adapters deteriorate the results of CodeBERT in most layers and do not improve the accuracy of the Graph-CodeBERT significantly (except a few layers). Another observation is the results of RoBERTa (dashed blue line), where it is consistently higher than the two C-PLMs. This indicates the ability of RoBERTa to predict the length of code. It could be related to the fact that though programming languages are different from natural language, this task is not necessarily differentiating the PL and NL, and thus the NL-PLM can still achieve good results for predicting the length of the given input.

{\textbf{Cyclomatic Complexity Results.}
The results of the cyclomatic complexity analysis are presented in Figure \ref{fig:cpx-prob}. The experiment results show that the performance of RoBERTa with language adapters is comparable to that of CodeBERT and GraphCodeBERT without adapters. Note that cyclomatic complexity is a highly complex task, which measures the amount of structural information that is encoded onto each Transformer block of a model.

It is worth noting that for all of the probing tasks considered in this study, the variations in accuracy follow a consistent trend. This is a strong indication that RoBERTa has been effectively adapted to programming languages, as \textit{RoBERTa+Adapters} exhibit similar behavior across all of these probing tasks.
This consistency in the results suggests that RoBERTa with language adapters has a good understanding of the code structure, and in general, it is well-suited for modeling programming languages.

\begin{figure}

\includegraphics[width=\columnwidth]{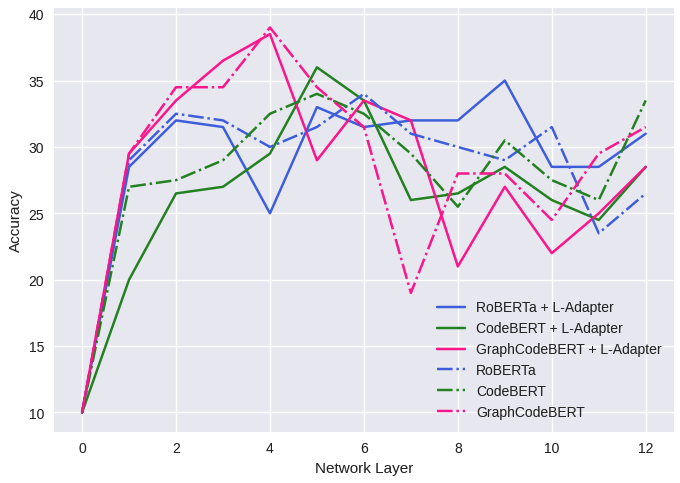}
\centering
\caption{The accuracy of predicting cyclomatic complexity in the probing task is evaluated on each layer of the model. The x-axis represents the classification results at each layer, with the first layer being the input embeddings, which serves as the naive baseline accuracy. All models utilized in the experiment have 12 transformer layers, and the y-axis displays the accuracy at each layer for each model.}
\label{fig:cpx-prob}
\end{figure}

\textbf{AST Node Tagging Results.}
Figure \ref{fig:type-probe} shows the results of AST node tagging for RoBERTa, CodeBERT, and GraphCodeBERT with and without language adapters. \textit{RoBERTa+L-Adapter} exhibits a lower accuracy compared to C-PLMs in the early layers (i.e., layers 3, 4, and 5). However, starting from the sixth layer, its performance becomes comparable to that of C-PLMs and in some cases (e.g., layers 10, 11, 12), it even outperforms GraphCodeBERT. 
It is important to note that not all code characteristics are expected to be improved in the subsequent layers of a model, and in the context of various models, it is noteworthy that their corresponding layers may exhibit varying performance levels when subjected to the same probing task.
In fact, certain characteristics may be specific to certain layers of the model. As seen in the baseline C-PLMs, there are some variations in accuracy across different layers.

\begin{figure}

\includegraphics[width=\columnwidth]{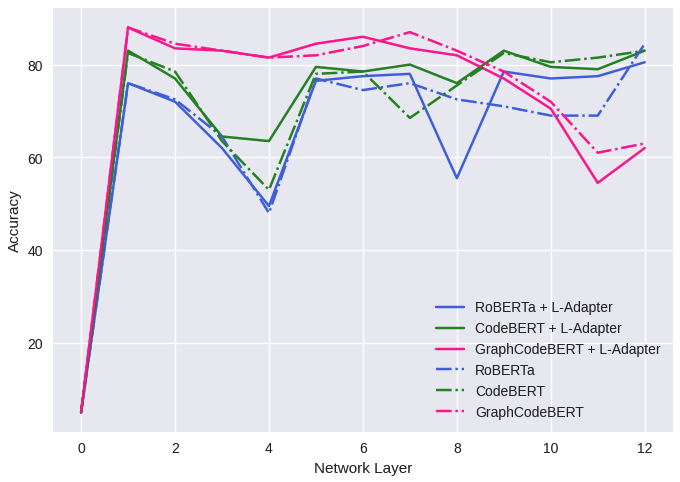}
\centering
\caption{Accuracies of AST node tagging probing task. The x-axis demonstrates the classification results at each layer (the first layer is the input embeddings, which represent the naive baseline accuracy). All models have 12 Transformer layers. The y-axis shows the accuracy at each layer for each model.}
\label{fig:type-probe}
\end{figure}

Note that we observe a similar pattern in Figure \ref{fig:type-probe} as in Figures \ref{fig:len-prob} and \ref{fig:cpx-prob}, for the C-PLM models with/without adapters. The L-adapters decrease the performance of the C-PLMs slightly. This behavior is expected. The C-PLMs are pre-trained on programming languages, and have captured the PL-related knowledge. Therefore, adding an L-adapter trained on one programming language, interferes with their current learning. This also explains the rationale of our choice, where we added only task adapters to the C-PLMs in our experiments. 
However, the NL-PLM is not pre-trained on programming languages. Therefore, we first need to introduce them to the programming language, through L-adapter, and then apply a task adapter for a code-related task. That is the reason for adding both L-adapters and T-adapters to RoBERTa, when we want to use them for a code-related task.

Furthermore, as Figures \ref{fig:len-prob}--\ref{fig:type-probe} show, the models have different behavior with the learned embeddings in all levels (i.e., syntactic, structural, and semantic). The performances are different in various layers. However, all the findings are aligned with results obtained in the field of Natural Language Processing (NLP) \cite{jawahar2019does}; a prevailing pattern emerges where surface features are positioned in the lower layers, syntactic features occupy the middle layers, and higher layers encapsulate the profound semantic features. This is more evident in Figure \ref{fig:len-prob}.

\subsection{Attention Analysis} \label{sec:attention-analysis}

In this section, we analyze the attention of the function name tokens in two code samples, one written in Ruby and the other in Go. The goal is to gain insights into how adapters influence attention at different layers when they are inserted into the PLMs, as compared to when they are excluded from the PLMs.

\begin{figure}

\includegraphics[width=\columnwidth]{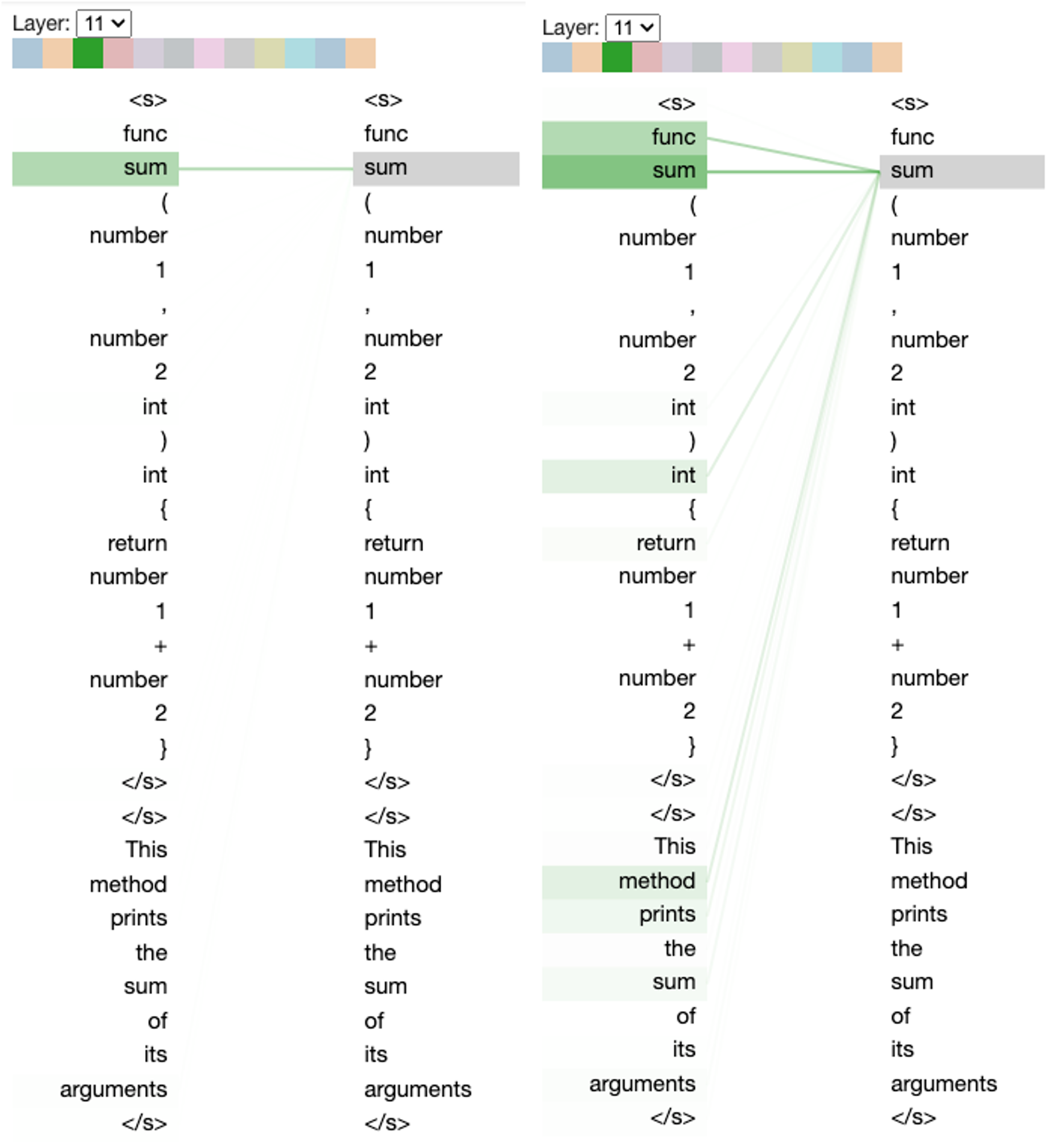}
\centering
\caption{An illustrative example of how adapters affect the last layer of RoBERTa when a Go sample is fed to the model. The left figure shows the attention of the third head on the function name \texttt{sum} without adapters whereas the right figure depicts the same attention head with adapters. As shown, while RoBERTa without adapters only attends to the local token neighbors, RoBERTa equipped with adapters has an in-depth knowledge of the code and pays more attention to the parts that are more related to the function name (e.g., it has strong attention to the \texttt{func} keyword which suggests that it knows that \texttt{sum} is somehow related to that keyword).}
\label{fig:go-attention}
\end{figure}

As an example, consider Figure \ref{fig:go-attention} which illustrates the attention behavior when a Go code sample is fed to RoBERTa. We use the Bertviz\footnote{https://github.com/jessevig/bertviz} tool to visualize the attention weights of the model.
On the left side, the attention of the last layer of RoBERTa is depicted without adapters, and it can be observed that the third head in the last layer only attends to the \verb|sum| token. 
The figure on the right side shows the same attention head when the adapters are inserted. 
On the right side, we observe that the same head in the same layer pays more attention to other tokens related to the function name when adapters are included. For example, the attention on the keyword \verb|func| indicates that the model bounds the function name with this keyword, or the tokens \verb|method|, \verb|prints|, \verb|the|, and \verb|sum| are semantically related to the function name, and so they are given more attention by the \verb|sum| token.

Another example can be seen in Figure \ref{fig:ruby-attention}, where a Ruby code sample is fed to RoBERTa, and the attention of the tenth layer from the fourth head is shown. On the left, adapters are excluded, and on the right, adapters are inserted into the NL-PLM, highlighting the impact of adapters on the attention gained by the model.
Similar observations are noticed in this figure too. When adapters are used, the attention of the token \verb|sum| to other related tokens increases.

Figures~\ref{fig:ruby-attention-codebert} and \ref{fig:ruby-attention-graphcodebert} demonstrate the effect of adding adapters to CodeBERT and GraphCodeBERT, respectively, in the same Ruby function, as the one used in Figure~\ref{fig:ruby-attention}. In Figure~\ref{fig:ruby-attention-codebert}, we observe that when adapters are added to CodeBERT, the model pays more attention to the parts related to the function name. The function name \texttt{sum} pays high attention to the \texttt{def}, \texttt{puts} and \texttt{+} keywords in the code, and the \texttt{prints}, \texttt{the} and \texttt{sum} words in the comment, which suggests that it can detect the boundaries of the function on this attention head. Figure~\ref{fig:ruby-attention-graphcodebert} presents another interesting observation. GraphCodeBERT and GraphCodeBERT with adapters attend to the same tokens more than other tokens in the code and comment; but when adapters are added, the intensity of the attention increases.

The provided examples suggest that adapters have a significant impact on the output of the internal embeddings. They operate by magnifying the embeddings in a way that the performance of the target task is optimized. This is an important aspect of fine-tuning PLMs, as it allows the model to adapt to specific tasks and improves its performance. The ability of the adapters to enhance the internal embeddings of the model is a key factor in its effectiveness for fine-tuning, and it is a powerful technique for improving the performance of PLMs on specific tasks.

\begin{figure}

\includegraphics[width=\columnwidth]{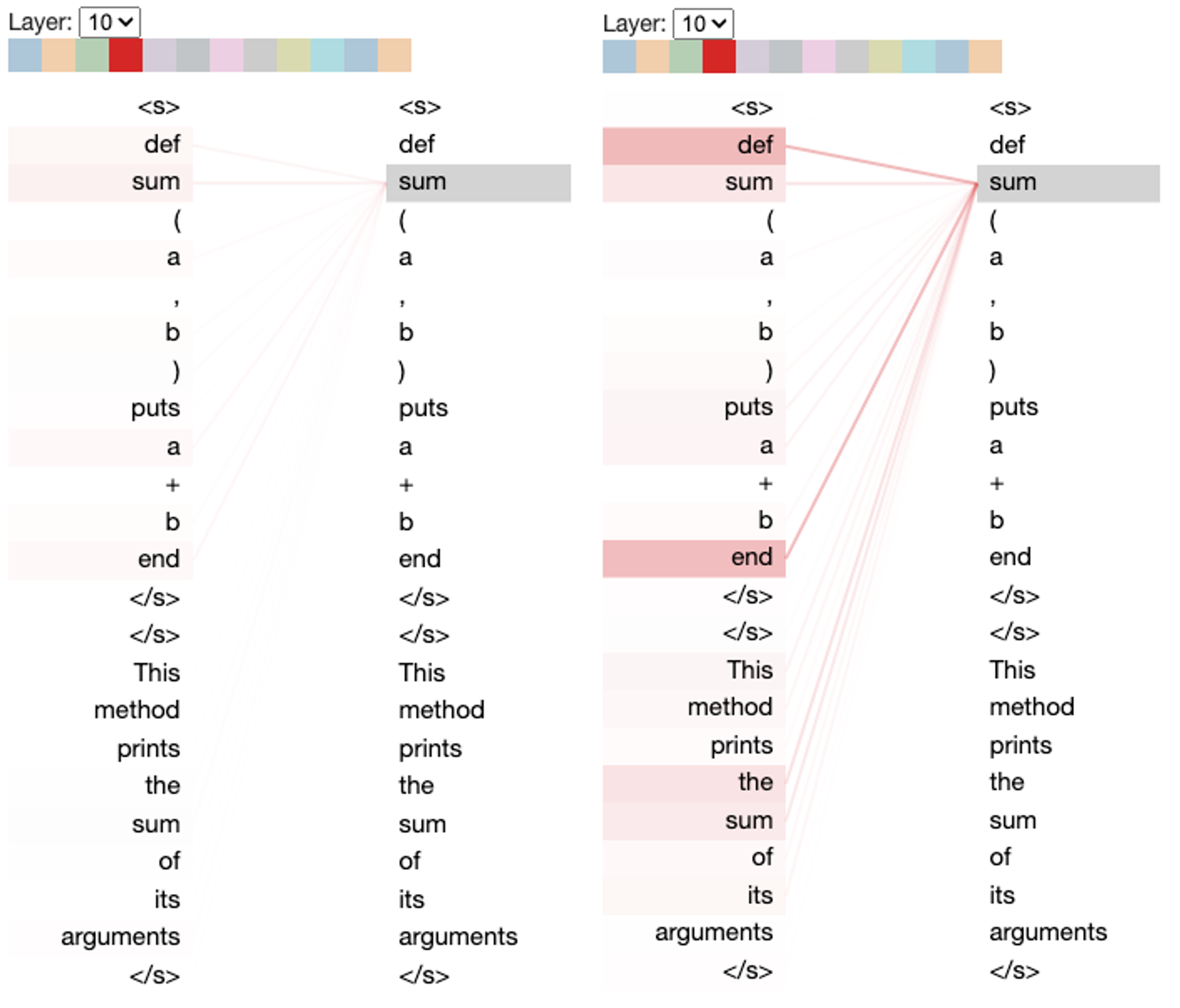}
\centering
\caption{The figures illustrate the tenth attention layer when a Ruby sample is fed to RoBERTa. The left figure shows the attention of the fourth head on the function name \texttt{sum} when adapters are not inserted. The right figure shows the same attention head when adapters are included. In this example, RoBERTa without adapters only has weak attention to the code tokens. On the other hand, RoBERTa equipped with adapters has stronger attention to the code and pays more attention to the parts more related to the function name (e.g., function name pays high attention to the \texttt{def} and \texttt{end} keywords which means it can detect the boundaries of the function on this attention head).}
\label{fig:ruby-attention}
\end{figure}

\begin{figure}

\includegraphics[width=\columnwidth]{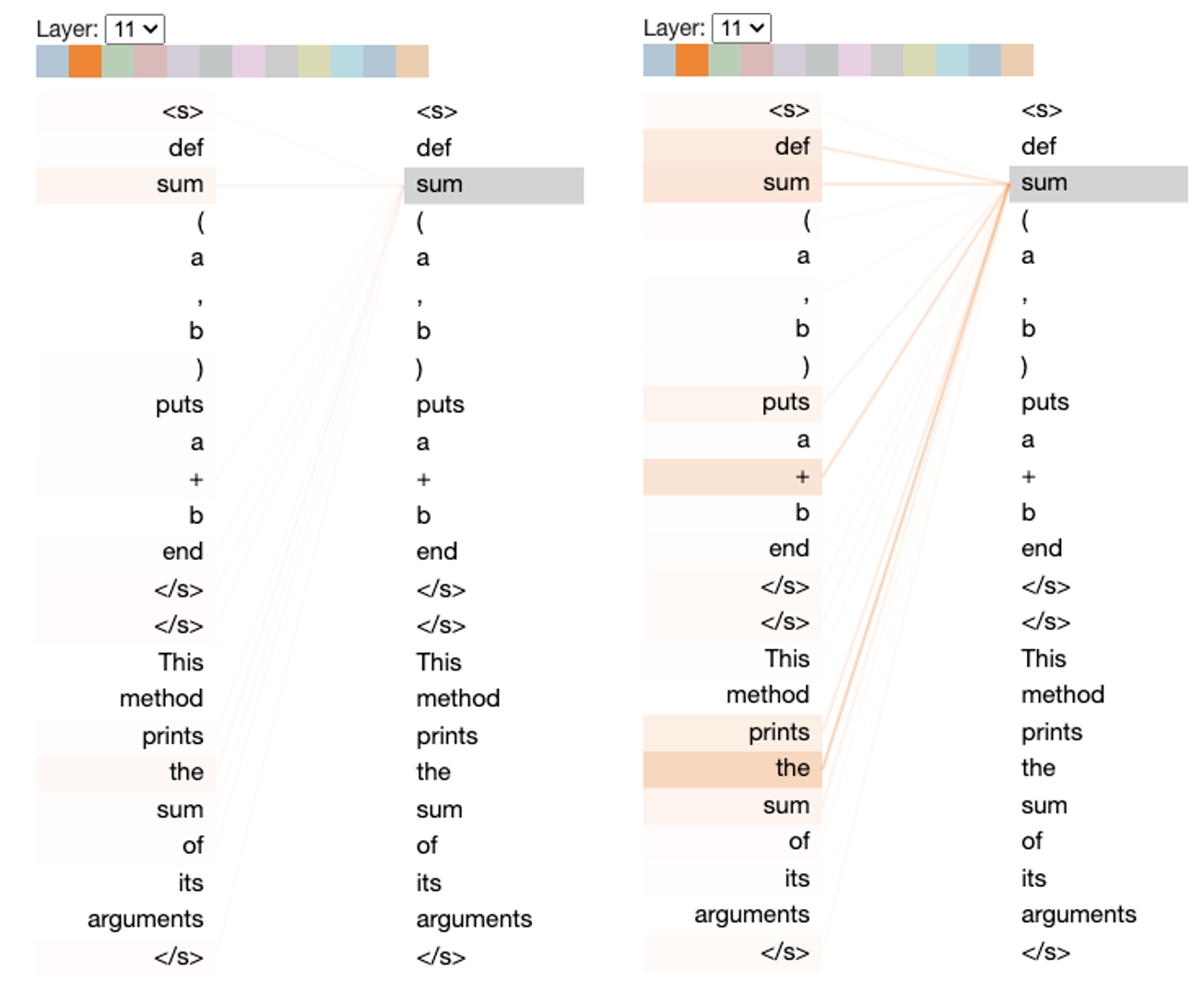}
\centering
\caption{The figures illustrate the eleventh attention layer when a Ruby sample is fed to CodeBERT model. The left figure shows the attention of the second head on the function name \texttt{sum} when adapters are not inserted. The right figure shows the same attention head when adapters are included. In this example, CodeBERT without adapters only has weak attention to the related tokens in code and comment. On the other hand, CodeBERT equipped with adapters has stronger attention to the code and pays more attention to the parts more related to the function name. Here, the function name \texttt{sum} pays high attention to \texttt{def}, \texttt{puts} and \texttt{+} keywords in the code, and \texttt{prints}, \texttt{the} and \texttt{sum} in the comment, which means it can detect the boundaries of the function on this attention head.}
\label{fig:ruby-attention-codebert}
\end{figure}

\begin{figure}

\includegraphics[width=\columnwidth]{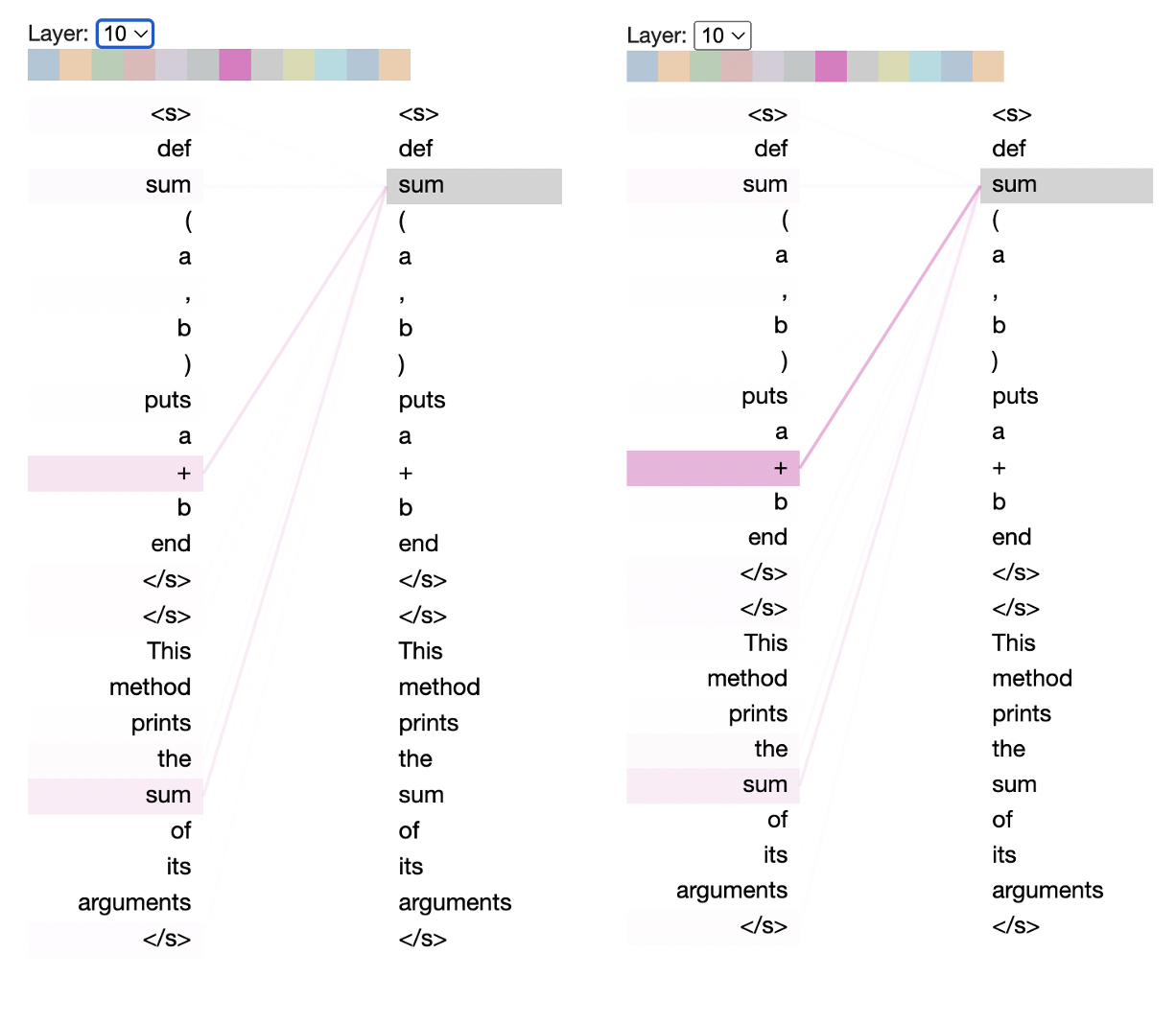}
\centering
\caption{The figures illustrate the tenth attention layer when a Ruby sample is fed to GraphCodeBERT model. The left figure shows the attention of the seventh head on the function name \texttt{sum} when adapters are not inserted. The right figure shows the same attention head when adapters are included. In this example, although both models pay attention to the same tokens, when we add adapters, it boosts the intensity of attention.}
\label{fig:ruby-attention-graphcodebert}
\end{figure}

\section{Implications} \label{sec:implications}

Fine-tuning PLMs for different tasks and using deep neural networks are known to be computationally expensive. Furthermore, not everyone has access to powerful resources such as high-end GPUs, which are essential to fine-tuning a PLM. On the other hand, an adapter is a pluggable module that can be inserted into a PLM, and the learned model requires less space than a fine-tuned PLM.
Adapters have been adopted in the NLP field recently, but their introduction for code representation and understanding is new. We anticipate different avenues that should be investigated in the SE domain, including searching for the best hyperparameters, identifying an optimized setup -- meaning that they should be inserted in all or some layers and how this affects their performance, and whether they can be used to integrate different tasks in SE. We note that these are some aspects for SE research, and they are not exhaustive. Multiple other research on SE-specific adapters could also be studied. 

We studied both NL-PLMs and C-PLMs with adapters. Interestingly, RoB-ERTa's results for some languages and some tasks were close to that of CodeBERT. This finding shows the potential of NL-PLM to be used for various code-related software engineering tasks. Researchers could conduct studies on the reasons behind such results and other ways to transfer this knowledge into the current C-PLMs or other code-specific trained models. 

In this study, we conducted experiments on three tasks, code clone detection, and code summarization, with a restricted number of programming languages. We observed that the results of using adapters in RoBERTa were better than the code clone detection results. Interestingly, the scores we obtained varied among different languages. Separate empirical studies could provide insights into the tasks or languages for which adapters are more useful.

Adapters are parameter and space efficient modules that enable scaling up a PLMs to multiple \textbf{tasks} and \textbf{languages}, without noting a significant drop in in-domain performance associated with the ``curse of multilinguality'' of a model \cite{conneau2019curseofMultilinguality}. 
The curse of multilinguality is related to PLMs trained on multiple languages, which is the trade-off between language coverage and model capacity. The limited capacity of a PLM leads to a performance drop when more languages are added to a multilingual model compared to its monolingual variants. However, the number of languages or tasks studied in NLP is much more than the ones used in the SE domain. Researchers can pursue this trade-off and find the capacity of using adapters for multiple languages and tasks when the number of programming languages used in pre-training a C-PLM increases. Developing more computation-efficient models and alternatives to fine-tuning models for code analysis is still in its initial steps for the SE domain, which could benefit from more research in this field.

Parameter efficient model may result in faster inference time. Due to the lower overhead of switching among fine-tuned PLMs, we can use the same NL-PLM (with adapters) or the same C-PLM (with adapters) for a higher number of tasks. This also allows us to better integrate the models in an Integrated Development Environment (IDE), as we will be integrating a single model instead of multiple fine-tuned models. Developing such a tool to help adapt the learned knowledge in NL-PLMs or C-PLMs and use it in an IDE could benefit a wider community.

\section{Threats to Validity} \label{sec:threats}

\textbf{External Validity} relates to the generalization of the results. This study is limited on the number of downstream tasks and the programming languages used. The results might not be generalizable to other tasks and programming languages. Though we expect similar results for other tasks, additional studies are required to validate this.


\textbf{Internal Validity} is related to having unanticipated relationships. 
One of the threats can be related to training the models. The authors who trained the models have extensive experience with adapter modules, and have both theoretical and technical knowledge on NLP and SE domains. Therefore, we anticipate minimal threats related to training the models. 

We used the CodeXGLUE benchmark, re-ran all the experiments for the PLMs, and confirmed the differences in the results that we obtained.
To mitigate obtaining unwanted results, we used the publicly available datasets from this benchmark platform and followed the steps mentioned in their pipeline to evaluate the models. 
Additionally, we conducted pilot studies to find the best setup for the adapters and baselines.


In our experiments, the selection of hyperparameters can impact the performance of the adapter fine-tuning phase. Determining the optimal values for these parameters is challenging and it is still an open research question. For example, the internal embeddings of adapters during down and upsampling are set to the default value, which could result in sub-optimal fine-tuning of the model. To mitigate this potential issue, we have used the hyperparameters suggested in \cite{pfeiffer2020madX}, as their work has conducted an extensive search for the best hyperparameters for adapters. However, the obtained results with adapters might still be improved with different hyperparameters.

\textbf{Construction Validity} relates to what a test claims to measure and what it actually measures. 
To mitigate such threats, we followed the evaluation metrics used for each task as it was done in previous studies. Additionally, for the probing analysis, we obtained the datasets and the tasks from previous studies 
 to ensure a consistent experimental setup.

\section{Conclusion and Future Works} \label{sec:conclusion-future}
In this paper, we studied the ability and efficiency of transferring the learned knowledge from an NL-PLM to programming language-based software engineering tasks through the use of adapters, thus assessing the ability of adapters to transfer from one modality to another modality.
Our results demonstrate that NL-PLMs equipped with adapters exhibit comparable performance to that of C-PLMs, suggesting that adapters can effectively adapt natural language models for programming-specific tasks such as code summarization and code clone detection. Additionally, we evaluated the utility of adapters as a rapid fine-tuning method for C-PLMs and found that even for models specifically designed for code, adapters can enhance performance on the code summarization task.
Training and fine-tuning adapters require a lower number of parameters with less storage.
Thus, adapters can be used in SE to scale up models for multiple tasks and languages, making them beneficial and practical in practice.
Adapters are pluggable modules, and they can be easily inserted for another language or task. 
We plan to study this characteristic and the training of multi-lingual adapters on code in the future. 

\section*{Data Availability Statement}

The data that support the findings of this study are available in the CodeXGLUE GitHub repository \footnote{https://github.com/microsoft/CodeXGLUE}.

\section*{Conflict of Interest}
The authors declare that they have no conflict of interest.

\section*{Acknowledgement}
We thank the original authors of the ICPC 2022 \cite{divyam} paper, Divyam Goel and Ramansh Grover. Part of the experiments from our previous paper coauthored with Divyam Goel and Ramansh Grover are rewritten in the current paper. \\This research is supported by a grant from the Natural Sciences and Engineering Research Council of Canada RGPIN-2019-05175.

\bibliographystyle{unsrt}
\bibliography{main}

\end{document}